\documentclass[aps,prb,groupedaddress,floatfix]{revtex4}

\usepackage{graphicx}
\usepackage{subfigure}

\def\be{\begin{equation}}
\def\ee{\end{equation}}
\def\bea{\begin{eqnarray}}
\def\eea{\end{eqnarray}}

\begin{document}


\title{Electronic instabilities of a Hubbard model approached as a large
array of coupled chains:
competition between $d$-wave superconductivity and pseudogap phase}


\author{E. Perfetto$^{1,2}$ and J. Gonz{\'a}lez$^{1}$}
\affiliation{$^{1}$Instituto de Estructura de la Materia.
        Consejo Superior de Investigaciones Cient{\'\i}ficas.
        Serrano 123, 28006 Madrid. Spain.\\
$^{2}$Consorzio Nazionale Interuniversitario per le Scienze
Fisiche della Materia, Universita' di Roma Tor Vergata, Via della
Ricerca Scientifica 1, 00133 Roma, Italy.}


\date{\today}

\begin{abstract}
We study the electronic instabilities in a 2D Hubbard model where one of
the dimensions has a finite width, so that it can be considered as  
a large array of coupled chains. The finite transverse size of the system
gives rise to a discrete string of Fermi points, with respective electron fields
that, due to their mutual interaction, acquire anomalous scaling dimensions
depending on the point of the string. Using bosonization methods, we show
that the anomalous scaling dimensions vanish when the number of coupled chains
goes to infinity, implying the Fermi liquid behavior of a 2D system in that limit.
However, when the Fermi
level is at the Van Hove singularity arising from the saddle points of the 2D
dispersion, backscattering and Cooper-pair scattering lead to the breakdown of
the metallic behavior at low energies. These interactions
are taken into account through their renormalization group scaling,
studying in turn their influence on the nonperturbative bosonization of the model.
We show that, at a certain low-energy scale, the anomalous electron
dimension diverges at the Fermi points closer to the saddle points of
the 2D dispersion. The $d$-wave superconducting correlations
become also large at low energies, but their growth is cut off as the
suppression of fermion excitations takes place first, extending progressively
along the Fermi points towards the diagonals of the 2D Brillouin zone. We stress
that this effect arises from the vanishing of the charge stiffness at the Fermi
points, characterizing a critical behavior that is well captured within our
nonperturbative approach.

\end{abstract}

\maketitle

\vspace*{1cm}
\section{Introduction}

Copper-oxide superconductors have shown since their discovery
quite unusual electronic properties, regarding the particular
features of the superconducting state as well as the behavior of
the normal state. One of the most remarkable experimental
observations has been the existence of the so-called pseudogap, at
temperatures which are above the superconducting dome in the phase
diagram\cite{tim}. The unconventional character of this pseudogap
lies in that it opens up quite anisotropically along the Fermi
line of the electron system. At some doping-dependent temperature
$T^*$, which is in general well above the transition temperature
of the superconducting phase, electron quasiparticles become
suppressed in the region of the Fermi line which is closer to the
momenta $(\pi, 0)$ and $(0, \pi)$. This is particularly manifest
in the ARPES experiments\cite{ding}, as the quasiparticle peaks
get a dispersion that does not cross the Fermi level at any point
in momentum space. As the temperature is lowered from $T^*$, this
partial destruction of the Fermi line takes place progressively
over wider regions, approaching the diagonals of the Brillouin
zone\cite{norman}.

There have been several attempts to provide a theoretical explanation for the
pseudogap in the cuprates. Some of the proposals have considered the possible
coupling of the electron quasiparticles to collective boson
excitations\cite{boson1,boson2}.
It has been also investigated the possibility that the electron system
may develop an ordered phase, which would induce the
appearance of a gap in the single-particle spectrum\cite{order1,order2}.
Other explanations are based on
the strong electron correlations and, in particular, on the description of the
cuprates in terms of the Hubbard model. Thus, it has been proposed that
theories based on a resonating-valence-bond state\cite{anderson} or on the
existence of states with time-reversal symmetry-breaking\cite{varma} may give
rise to features consistent with the pseudogap phase.

The main challenge that face all these proposals
is actually the difficulty to deal with a regime of strong electron
correlations. In these conditions, one has usually to resort to mean-field
approximations, that may give a qualitative idea of the possible phases of the
electron system. It would be also desirable to have an alternative
framework in which the nature of the pseudogap feature could be clarified.
In this regard, the motivation of the present paper is to investigate the
Hubbard model by means of a nonperturbative approach, trying to
capture the effects of the
strong electron correlations. For that purpose, we will consider a 2D model
where one of the dimensions has a finite width, so that it may be considered
as formed by the coupling of a large number of Hubbard chains.
A general analysis of the $N$-chain Hubbard model in weak coupling has been
accomplished in Ref. \onlinecite{lbf}.
We will use bosonization techniques to determine the nonperturbative effects of
the interactions with small momentum-transfer on the properties of the fermion
excitations. In this framework, the Fermi liquid properties of a 2D system are
recovered when the number of coupled chains goes to infinity, as the anomalous
scaling dimensions of the electron fields vanish in that limit.
The description has to be completed anyhow by considering the interactions
whose momentum-transfer is not small. These
can be taken into account through their renormalization group scaling,
studying their influence on the nonperturbative bosonization of the model and
the instabilities they may trigger at low energies.

The most interesting instance corresponds to the case where the Fermi level
is near the Van Hove singularity arising from the saddle points in the 2D
dispersion at $(\pi, 0)$ and $(0, \pi)$. When the 2D model is
approached by a large array of coupled chains, the
fermion excitations are attached to a large but finite number of Fermi points,
with respective electron fields that are
renormalized by the interactions and characterized by their anomalous
scaling dimensions. We will see that, for the model with
on-site repulsion, the angle-resolved quasiparticle weight vanishes below a
certain low-energy scale, starting from the Fermi points which are closer to
$(\pi, 0)$ and $(0, \pi)$. The $d$-wave superconducting correlations are also
dominant at low energies, but their growth is cut off as the suppression of
fermion excitations takes place first, extending progressively along the Fermi
points towards the diagonal of the Brillouin zone.

The results that we present are in part reminiscent of those obtained in
2D electron systems when the Fermi line is placed close to the
saddle points at $(\pi, 0)$ and
$(0, \pi)$\cite{dwave,eur,iof,schu,nucl,jap,metz,prl,hon,zan,binz,kat,kat2}.
It has been shown that
the presence of a Van Hove singularity induces a strong renormalization of the
quasiparticle properties, with a clear suppression of the quasiparticle weight
near the saddle points\cite{nucl,kat2}.
Some studies have also remarked that the divergent
flow of the couplings observed in renormalization group analyses could lead
to the vanishing of both the compressibility and the magnetic
susceptibility\cite{metz,hon,zan}.
In our approach, we are able to identify the onset of the destruction of the
Fermi line at an energy scale where the electron correlations have not
entered yet the divergent regime. This is possible as long as the Fermi surface
instability appears as a critical point in the nonperturbative quasiparticle
renormalization from the bosonization approach.
The present combination of bosonization and renormalization group methods may be
seen as an alternative to conventional analyses of 2D systems near the Van Hove
filling, offering a sensible 2D limit as the number of coupled chains is
increased.

\section{Quasiparticle properties in the limit of large number
of chains}

We will take as our starting point a 2D model with nearest-neighbor and 
next-to-nearest-neighbor hopping, with a finite size in one of the spatial 
dimensions. For the time being, we introduce a generic
spin-independent interaction with potential $V$. The hamiltonian of the
system is then given by
\begin{eqnarray}
H   & = &  \sum_{j=1}^{N} \sum_{n=-\infty}^{\infty} 
         \left( -t \; \psi^{\dagger }_{j \sigma} (n) \psi_{j \sigma} (n+1)
                -t \; \psi^{\dagger }_{j \sigma} (n) \psi_{j+1  \sigma} (n)
                   -t' \psi^{\dagger }_{j \sigma} (n) \psi_{j+1  \sigma} (n \pm 1)
                           + {\rm h. c.}   \right)           \nonumber    \\
   &   &   +  \sum_{j,l=1}^{N} \sum_{m,n=-\infty}^{\infty}
      \psi^{\dagger }_{j \sigma} (m) \psi_{j \sigma} (m)  \: V_{jl} (m, n) \:
                   \psi^{\dagger }_{l \sigma'} (n) \psi_{l \sigma'  } (n)
\label{ham}
\end{eqnarray}
where $\psi^{\dagger }_{j \sigma} (n)$ and $\psi_{j \sigma} (n)$  are electron creation and
annihilation operators at site $n$ of the $j$-th chain, with spin projection
given by the $\sigma $ index.

\begin{figure}[h]
\includegraphics[height=4cm ]{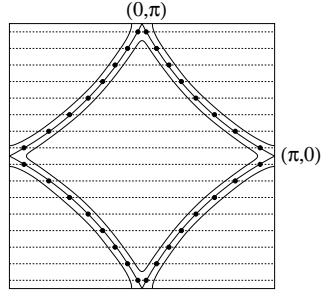}
\caption{Schematic representation of the Fermi points for a 2D model made of a
finite number of coupled chains, when the Fermi level is at the Van
Hove singularity. The horizontal lines correspond to quantized transverse
momenta for twisted boundary conditions, $k_a =  \pi (2a+1)/N $, with
$a = -N/2, -N/2 + 1, \ldots N/2 - 1$. Energy contour lines are also represented
for the cutoff at $E_c$ and $-E_c$ about the Fermi level.}
\label{one}
\end{figure}

For the description of quasiparticle properties, it is convenient to make the
passage to electron operators in momentum space, taking the Fourier transform
in the site and chain variables:
\begin{equation}
\psi_{j \sigma} (n) =  \frac{1}{\sqrt{N}} \sum_{a} \int^{\pi}_{-\pi} dp
             \;  e^{-i k_a j} e^{-i p n}  \psi_{\sigma} (k_a, p)
\end{equation}
We will apply periodic boundary conditions to the system of $N$ chains.
Whenever it is convenient to exclude the singular points $(\pi , 0)$ and
$(0, \pi )$ from the spectrum, we will make however a twist by $e^{i\pi }$ in
the electron field after completing a whole period along the array. With
these twisted boundary conditions, the transverse momentum $k_a$ will be
quantized according to the rule
$k_a =  \pi (2a+1)/N, \; a = -N/2, -N/2 + 1, \ldots N/2 - 1$.
From the hamiltonian (\ref{ham}), we get the dispersion relation for the
different transverse momenta
\begin{equation}
\varepsilon_{a} (p) = -2t \cos (p) - 2t \cos (k_a)
                        -4t' \cos (p) \cos (k_a)
\end{equation}
Furthermore, for the sake of describing low-energy properties, we may
concentrate on the modes that are within a cutoff energy $E_c $
about the Fermi energy $\varepsilon_F$, where the dispersion becomes
approximately linear about the Fermi points. This is illustrated in
Fig. \ref{one}. For each transverse momentum
$k_a$, the low-energy excitations can be encoded into two fields
$\psi_{a \sigma} (p)$ and $\psi_{\overline{a} \sigma} (p)$ with opposite
chirality (i.e. right- or left-moving character), corresponding to the two
different Fermi points at $p_a$ and $-p_a$ :
\begin{eqnarray}
\lefteqn{ \int dp \; \psi^{\dagger }_{\sigma} (k_a, p) \left( \varepsilon_{a} (p) -
   \varepsilon_F  \right) \psi_{\sigma} (k_a, p)  }        \nonumber            \\
    &  \approx  &    \int dp  \;
      v_a   \left( \psi^{\dagger }_{a \sigma} (p) ( p - p_a ) \psi_{a \sigma} (p)
     +   \psi^{\dagger }_{\overline{a} \sigma} (p)
                          (- p - p_a ) \psi_{\overline{a} \sigma} (p)    \right)
\end{eqnarray}
$v_a$ being the Fermi velocity at $p_a$ and $-p_a$.

From the detailed discussion in Ref. \onlinecite{lbf}, it follows that
$E_c $ can be taken suitably small so that the possible scattering processes
are severely limited by momentum conservation. The price that one has to pay
for the reduction of the cutoff is that the original interactions are
corrected by terms which are of the order of
$\sim (V/2\pi t)^2 \log (k_F / E_c)$, $k_F$ being the scale of the size of
the Fermi line. In our model, we will start with an on-site interaction $U$
such that $U/2 \pi t < 1$, in such a way that the perturbative corrections
are under control down to a conveniently small cutoff $E_c$.

\begin{figure}[h]
\includegraphics[height=8cm ]{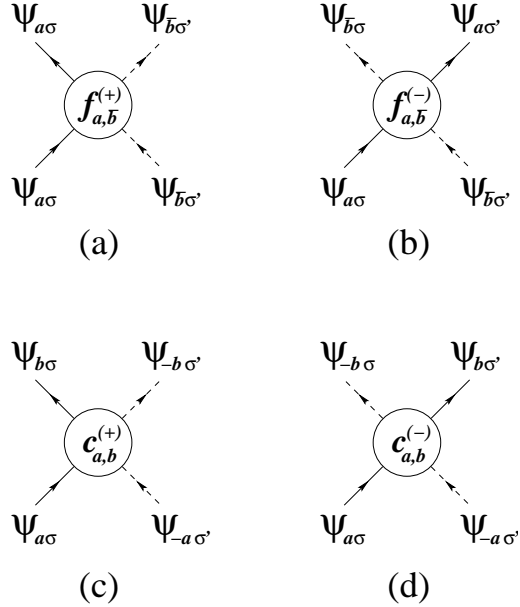}
\caption{Interaction processes that survive at low energies within the cutoff
$E_c$ about the Fermi level. The solid and dashed lines correspond to 
the two different chiralities (right- and left-moving character) of the electron 
fields.}
\label{two}
\end{figure}

Confining our discussion to the low-energy modes in the interval between
$\varepsilon_F - E_c $ and $\varepsilon_F + E_c $, 
there is a reduced number of ways in which the momentum
can be conserved in the scattering processes. These can be classified into four 
different interaction
channels, as represented in Fig. \ref{two}. We may have processes
with suitably small momentum transfer between interacting particles near Fermi
points $a$ and $b$ (to which we will assign couplings $f^{(+)}_{a,b}$) and
processes where the particles exchange their positions in the neighborhood of
the two Fermi points (which we will label with couplings $f^{(-)}_{a,b}$).
The other type of interactions that exist generically at low energies
corresponds to processes where the total momentum of the incoming particles is
almost vanishing. In this case, we can discern a channel where the particles
interact keeping their respective chiralities (with couplings that we will
denote by $c^{(+)}_{a,b}$) and another where they exchange their chiralities
(requiring then different couplings $c^{(-)}_{a,b}$) \cite{foot}. The
hamiltonian involving the low-energy excitations becomes then
\begin{eqnarray}
H  & = &  \sum_a \int dp  \;
      v_a   \left( \psi^{\dagger }_{a \sigma} (p) ( p - p_a ) \psi_{a \sigma} (p)
 + \psi^{\dagger }_{\overline{a} \sigma} (p) (- p - p_a ) \psi_{\overline{a} \sigma} (p)
                                        \right)       \nonumber       \\
  &    &   +   \sum_{<a,b>}  \int dp dp' dq
  \left[  \left(  
  \psi^{\dagger }_{a \sigma} (p+q) \psi_{a \sigma} (p) \; f^{(+)}_{a,b} \;
             \psi^{\dagger }_{b \sigma'} (p'-q) \psi_{b \sigma'} (p')
                                          \right.  \right.    \nonumber    \\
  &  &  +  \psi^{\dagger }_{b \sigma} (p+q) \psi_{a \sigma} (p)  \; f^{(-)}_{a,b} \;
             \psi^{\dagger }_{a \sigma'} (p'-q) \psi_{b \sigma'} (p')      \nonumber    \\
  &  &  +   \psi^{\dagger }_{a \sigma} (p+q) \psi_{a \sigma} (p)
                                               \; f^{(+)}_{a,\overline{b}} \;
             \psi^{\dagger }_{\overline{b} \sigma'} (p'-q) \psi_{\overline{b} \sigma'} (p')
                                                      \nonumber    \\
  &  &  \left.  +  \psi^{\dagger }_{\overline{b} \sigma} (p+q) \psi_{a \sigma} (p)
                                              \; f^{(-)}_{a,\overline{b}} \;
             \psi^{\dagger }_{a \sigma'} (p'-q) \psi_{\overline{b} \sigma'} (p') 
  +  a \leftrightarrow \overline{a} , b \leftrightarrow \overline{b} \right) \nonumber    \\
  &  &  +  \psi^{\dagger }_{b \sigma} (p+q) \psi_{a \sigma} (p)  \; c^{(+)}_{a,b} \;
             \psi^{\dagger }_{-b \sigma'} (p'-q) \psi_{-a \sigma'} (p')      \nonumber    \\
  &  &   \left.   +  \psi^{\dagger }_{-b \sigma} (p+q) \psi_{a \sigma} (p)  \;
         c^{(-)}_{a,b} \;
             \psi^{\dagger }_{b \sigma'} (p'-q) \psi_{-a \sigma'} (p')    \right]
\label{hle}
\end{eqnarray}
where the indices in the sum over pairs $<a,b>$ run over just one
of the chiralities. 

To clarify our notation, we recall that the index $\overline{a}$ corresponds 
to a Fermi point with the same transverse momentum but opposite chirality to 
that represented by the index $a$. Thus, the difference between the 
interactions $f^{(+)}_{a,b}$ and $f^{(+)}_{a,\overline{b}}$ is that the first
involves modes which have the same chirality, while  $f^{(+)}_{a,\overline{b}}$
couples an electron current made of right-moving fields to another current of
left-moving fields. On the other hand, the difference between the interactions
represented by $f^{(+)}_{a,\overline{b}}$ and $f^{(-)}_{a,\overline{b}}$ is 
that, in the latter, the incoming particles exchange their chirality in the 
scattering process. We finally remark that, in our notation, the index $-a$ 
represents a Fermi point with a momentum completely opposite to that of Fermi 
point $a$ (so that the modes around Fermi point $-a$ have opposite
chirality to those around Fermi point $a$). In this way, the incoming particles
are bound to have very small total momentum in the interactions represented 
by $c^{(+)}_{a,b}$ and $c^{(-)}_{a,b}$ \cite{note}.

An important difference between the interactions of particles with like
chirality ($f^{(+)}_{a,b}, f^{(+)}_{\overline{a},\overline{b}}, 
f^{(-)}_{a,b}, f^{(-)}_{\overline{a},\overline{b}}$) and those corresponding 
to $f^{(+)}_{a,\overline{b}}, f^{(-)}_{a,\overline{b}}, c^{(+)}_{a,b},
c^{(-)}_{a,b}$ is that the strength of the latter depends on the energy scale.
This may induce instabilities that appear when the system  is observed at
scales very close to the Fermi level. On the other hand, the significance of
the interactions with small momentum transfer
($f^{(+)}_{a,b}, f^{(+)}_{\overline{a},\overline{b}}, f^{(+)}_{a,\overline{b}},
f^{(+)}_{\overline{a}, b}$) 
is that they dictate the class of electron liquid describing the normal state
(metallic) properties of the coupled chains. This corresponds to the so-called
Luttinger liquid for any finite number of chains\cite{sol}. We discuss next
the way in which the limit of a large number of chains may change the
quasiparticle properties, depending on the given Fermi point.

The Luttinger liquid properties are best described in terms of operators
given by the density fluctuations at small momentum:
\begin{eqnarray}
B^{\dagger }_a (q)  & = &  \frac{1}{\sqrt{|q|}}
           \int dp \; \psi^{\dagger }_{a \sigma} (p+q)  \psi_{a \sigma} (p)
                                 \;\;\;\;\;\;  q > 0     \label{b1}       \\
B^{\dagger }_{\overline{a}} (q)  & = &  \frac{1}{\sqrt{|q|}}
  \int dp \; \psi^{\dagger }_{\overline{a} \sigma} (p+q)  \psi_{\overline{a} \sigma} (p)
                                   \;\;\;\;\;\;  q < 0
\label{b2}
\end{eqnarray}
It is well-known that the kinetic part of the hamiltonian (\ref{hle}) can be
written in terms of the operators (\ref{b1}) and (\ref{b2}) \cite{sol}. For the
time being, we will restrict the discussion to the interactions preserving the
chirality of the particles. Then, we end up with the hamiltonian governing
the Luttinger liquid dynamics
\begin{eqnarray}
H_{LL}  & = &  \sum_a v_a  \left( \int_{q > 0} dq \:|q|\: B^{\dagger }_a (q) B_a (q)
  + \int_{q < 0} dq \:|q|\: B^{\dagger }_{\overline{a}} (q) B_{\overline{a}} (q)
                                        \right)        \nonumber        \\
   &   &  + \sum_{<a,b>} \int_{q > 0}  dq |q|
     \left(  f^{(+)}_{a,b} ( B^{\dagger }_a (q) B_b (q) + B_a (q) B^{\dagger }_b (q) )
      +  f^{(+)}_{a,\overline{b}} ( B^{\dagger }_a (q) B^{\dagger }_{\overline{b}} (-q)
         + B_a (q) B_{\overline{b}} (-q) )      \right)   \nonumber     \\
   &   &  + \sum_{<a,b>} \int_{q < 0}  dq |q|
     \left(  f^{(+)}_{\overline{a}, \overline{b}}
   ( B^{\dagger }_{\overline{a}} (q) B_{\overline{b}} (q)
           + B_{\overline{a}} (q) B^{\dagger }_{\overline{b}} (q) )
      +  f^{(+)}_{\overline{a}, b} ( B^{\dagger }_{\overline{a}} (q) B^{\dagger }_b (-q)
         + B_{\overline{a}} (q) B_b (-q) )      \right)
\label{hll}
\end{eqnarray}

The hamiltonian $H_{LL}$ can be diagonalized by making the change of variables
\begin{eqnarray}
\widetilde{B}^{\dagger }_a (q) & = &  \sum_b   \left( s_{ab} B^{\dagger }_b (q) +
                   t_{a \overline{b}} B_{\overline{b}} (-q)   \right)      \\
\widetilde{B}_{\overline{a}} (-q) & = &  \sum_b
         \left( s_{\overline{a} \overline{b}} B_{\overline{b}} (-q)
           +  t_{\overline{a} b} B^{\dagger }_b (q)   \right)
\label{trans}
\end{eqnarray}
The change of basis defined by the matrix elements $s_{ab}, t_{a \overline{b}}$
is not unitary, since the requirement of canonical commutation relations for
the new operators imply in particular that
\begin{equation}
\sum_b ( |s_{ab}|^2 - |t_{a \overline{b}}|^2 ) = 1
\end{equation}
For the sake of carrying out the diagonalization of $H_{LL}$ with great
computational efficiency, one can introduce however the following trick. By
redefining all the boson operators with left-handed character, for instance,
\begin{equation}
B_{\overline{a}} (q) = i \widehat{B}_{\overline{a}} (q)   \;\;\;\;\; ,
 \;\;\;\;\;   B^{\dagger }_{\overline{a}} (q) = i \widehat{B}^{\dagger }_{\overline{a}} (q)
\label{redef}
\end{equation}
we change the canonical commutation relations to
\begin{equation}
\left[ \widehat{B}_{\overline{a}} (q) ,
            \widehat{B}^{\dagger }_{\overline{a}} (q')  \right] = - \delta (q - q')
\end{equation}
Keeping $B_{a} (q)=\widehat{B}_{a} (q), B^{\dagger }_{a} (q)=\widehat{B}^{\dagger }_{a} (q)$
for the other chirality, the linear change of variables for the redefined
operators has matrix elements that correspond to a unitary change of basis.
Thus, by performing first the redefinition (\ref{redef}), one can apply then
standard algorithms that are able to diagonalize very large complex matrices.
By reverting at the end the transformation (\ref{redef}), it is possible then
to bring the hamiltonian (\ref{hll}) to the form
\begin{equation}
H_{LL} = \sum_a \widetilde{v}_a \left(\int_{q > 0} dq |q| \widetilde{B}^{\dagger }_a (q)
  \widetilde{B}_a (q)  + \int_{q < 0} dq |q| \widetilde{B}^{\dagger }_{\overline{a}} (q)
   \widetilde{B}_{\overline{a}} (q)     \right)
\label{hd}
\end{equation}

Our aim is to show the evolution of the quasiparticle properties at the
hot spots around the saddle points $(\pi , 0)$ and $(0 , \pi)$. 
As a most relevant instance, we give the results of performing the 
diagonalization of the hamiltonian (\ref{hll}) for a growing number of 
coupled chains with $t' = -0.2 t$, at the 
filling that makes the 2D Fermi line to go across the saddle points. We have 
taken an on-site interaction $U$, that leads to a set of momentum-independent 
couplings equal to $U/N$ in the hamiltonian (\ref{hle}). In general we may 
interpret that the original boson operators are dressed by the interactions. 
The different correlation functions can be obtained by expressing the 
operators $B^{\dagger }_{a} (q)$ and $B_{a} (q)$ in terms of the set of 
noninteracting operators in (\ref{hd}), in such a way that the interaction 
effects are encoded in the string $\{ \widetilde{v}_a\}$ and in the parameters 
of the transformation (\ref{trans}).


The most important piece of information comes, after the diagonalization, from
the change of basis that brings the hamiltonian to the form (\ref{hd}).
The original boson operators can be obtained from the free operators that appear
in (\ref{hd}) through the inverse change of variables
\begin{eqnarray}
B^{\dagger }_a (q) & = &  \sum_b   \left( \tilde{s}_{ab} \widetilde{B}^{\dagger }_b (q) +
  \tilde{t}_{a \overline{b}} \widetilde{B}_{\overline{b}} (-q)   \right)   
\label{cv1}                                                                \\
B_{\overline{a}} (-q) & = &  \sum_b
 \left( \tilde{s}_{\overline{a} \overline{b}} \widetilde{B}_{\overline{b}} (-q)
           +  \tilde{t}_{\overline{a} b} \widetilde{B}^{\dagger }_b (q)   \right)
\label{cv2}
\end{eqnarray}
In the absence of interaction, the only nonvanishing coefficients are the 
diagonal elements $\tilde{s}_{aa} = \tilde{s}_{\overline{a} \overline{a}} = 1$.
The interactions give rise however to anomalous scaling dimensions $\gamma_a$
of the electron fields, which can be obtained from the
spatial decay of the electron propagator at equal time $t$
\begin{equation}
\langle \psi^{\dagger }_{a \sigma} (x,t)  \psi_{a \sigma} (y,t) \rangle
       \sim  \frac{1}{| x - y |^{1 + \gamma_a }}
\end{equation}
It is well-known that the fermion fields $\psi_{a \sigma} ,
\psi_{\overline{a} \sigma}$ are recovered by exponentiation of the integral
of the corresponding chiral part of the uniform density $\rho_a (x)$ \cite{sol}.
This is given according to (\ref{b1}) and (\ref{b2}) by
\begin{equation}
\rho_a (x) \approx \int_{q > 0} dq \sqrt{|q|} \; e^{-iqx}
     \left( B^{\dagger }_{a} (q) +  B_{\overline{a}} (-q)  \right)
      +   \int_{q < 0} dq \sqrt{|q|} \; e^{-iqx}
     \left( B^{\dagger }_{\overline{a}} (q) +  B_a (-q)  \right)
\label{dens}
\end{equation}
Then, it can be seen that
\begin{equation}
1 + \gamma_a =
 \sum_b \left( |\tilde{s}_{ab}|^2 + |\tilde{t}_{a \overline{b}}|^2 \right)
\label{anmd}
\end{equation}
We observe that, as soon as the interaction is switched on, we get an 
anomalous dimension $\gamma_a \neq 0$, reflecting the disappearance of the 
quasiparticle pole in the electron propagator.

\begin{figure}
\includegraphics[height=4.0cm ]{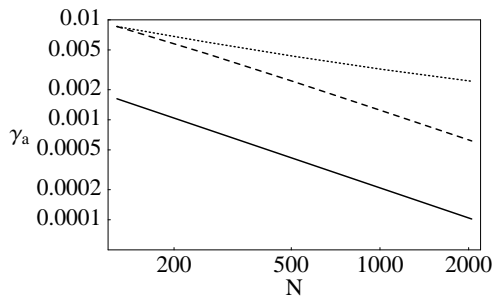}
\caption{Dependence of the anomalous electron dimension $\gamma_a$ on the
number $N$ of coupled chains in a Hubbard model with periodic boundary
conditions, when the Fermi point is taken close to the diagonal of
the Brillouin zone (full line), with a constant transverse momentum
$k_a = \pi /64$ (dashed line), and at the nearest location to the saddle point
with $k_a = 2\pi /N$ (dotted line).}
\label{four}
\end{figure}

It is very instructive to observe the behavior of the anomalous dimension
$\gamma_a$ at different Fermi points of the model. We find that, away from
$(\pi , 0)$ and
$(0, \pi )$, the anomalous dimension $\gamma_a$ decreases in the limit of
large number $N$ of chains, following with great accuracy a $1/N$
behavior\cite{kop}.
The prefactor for this law depends however on the distance to the saddle
points, and it grows as one gets closer to $(\pi , 0)$ or $(0, \pi )$, as
shown in Fig. \ref{four}. The limit towards these points becomes actually
quite singular. If one takes the limit of large $N$ sitting at the Fermi point
which is away but nearest to the saddle point (so that its Fermi velocity
decreases consequently as $1/N$) we observe that the anomalous dimension also
decreases monotonically, but with a behavior that seems to be smoother than
any power-law, as observed from Fig. \ref{four}. The sizes for which we have
been able to diagonalize the model (up to $\approx 2000$ chains)
do not allow us to assure whether this
limit to the saddle point may produce a vanishing anomalous dimension as
$N \rightarrow \infty$. Our results make clear anyhow that, for a Fermi point
placed at any finite distance from the saddle points, the anomalous dimension
vanishes in the limit of very large number of chains. This is what allows to
take the limit $N \rightarrow \infty$ as a way of approaching the properties
of an interacting Fermi liquid in two dimensions. We will see actually that
the presence of backscattering and Cooper-pair scattering modify significantly
the present picture, leading to an unconventional low-energy behavior near the
hot spots for large number of Fermi points.

\section{Renormalization group approach to low-energy phases}

In order to describe the quasiparticle properties at low energies, one has
to incorporate the effects of the variation of the couplings
$f^{(+)}_{a,\overline{b}}, f^{(-)}_{a,\overline{b}}, c^{(+)}_{a,b},
c^{(-)}_{a,b}$ with the energy scale. The discussion of the preceding section
is pertinent at the energy scale $E_c$ where the dispersion becomes
approximately linear about the Fermi level. At lower energy scales, the
effective values of the couplings may change significantly, having a large
impact on the renormalization of the quasiparticle properties.

We have to remark at this point that the renormalization effects manifest 
in general not only in the quasiparticle parameters,
but also giving rise to changes in the shape of the Fermi surface. It has
been proposed that, within a renormalization group approach at constant
particle number, the Fermi surface has to be understood as a fixed point of 
the renormalization group\cite{lk}. A practical implementation of this 
idea for a two-chain model can be found in Ref. \onlinecite{lkf}. In the present 
context, to find such a fixed-point solution is beyond reach 
for the large string of Fermi points, as the Fermi surface is reshaped also 
by irrelevant interactions whose momenta are not constrained by the cutoff, 
and that proliferate at large $N$ even more rapidly than the couplings 
already considered. Being interested in the situation where the Fermi level 
is close to a Van Hove singularity, it will be enough for us to guarantee 
that the Fermi line is not driven away from the saddle points at $(\pi, 0)$ 
and $(0, \pi)$ upon switching the interaction. 
We rely for that on the evidence obtained with the renormalization group 
at fixed chemical potential, which provides a sensible approach to systems 
in contact with a large reservoir of particles. In this context, it has been 
actually shown that the Van Hove singularity acts as an attractor, pinning 
the Fermi level for a certain window in the choice of the chemical 
potential\cite{eur,nucl}. We motivate in this way the suitability of 
studying the model with the Fermi line near $(\pi, 0)$ and $(0, \pi)$, as 
this choice does not imply the fine-tuning of the chemical 
potential in the system (as fixed for instance by the contact with a charge 
reservoir).

Below the scale $E_c $, the four-fermion interaction vertices are
corrected by diagrams that depend logarithmically on the energy scale
$\Lambda $ (measured from the Fermi energy $\varepsilon_F $). To deal with
this divergence, one can integrate progressively the electron modes starting
from the limits of the linear branches at $\varepsilon_F \pm E_c $ \cite{shankar}.
This leads to a dependence of the couplings on the scale
$l = \log (E_c /\Lambda )$, governed by scaling equations that read
at the one-loop level
\begin{eqnarray}
\frac{\partial f^{(+)}_{a,\overline{b}} }{ \partial l}  &  =  &
  -\frac{1}{2\pi v_{ab}} (  (f^{(-)}_{a,\overline{b}})^2
                            -  (c^{(+)}_{a,-\overline{b}})^2  )
                                            \label{fs1}          \\
\frac{\partial f^{(-)}_{a,\overline{b}} }{ \partial l}  &  =  &
   -\frac{1}{\pi v_{ab}} (  (f^{(-)}_{a,\overline{b}})^2
   +  (c^{(-)}_{a,-\overline{b}})^2
    -  c^{(-)}_{a,-\overline{b}}  c^{(+)}_{a,-\overline{b}}  )   \\
\frac{\partial c^{(+)}_{a,b} }{ \partial l}  &  =  &
   -\sum_{c}   \frac{1}{2\pi v_c}
      ( c^{(+)}_{a,c} c^{(+)}_{c,b} + c^{(-)}_{a,c} c^{(-)}_{c,b} )
    +  \frac{1}{2\pi v_{ab}}   c^{(+)}_{a,b} h^{(+)}_{a,-b}
                                                 \label{coop1}      \\
\frac{\partial c^{(-)}_{a,b} }{ \partial l}  &  =  &
   -\sum_{c}    \frac{1}{2\pi v_c}
     ( c^{(+)}_{a,c} c^{(-)}_{c,b} + c^{(-)}_{a,c} c^{(+)}_{c,b} )
                                                        \nonumber         \\
  &  &     -  \frac{1}{\pi v_{ab}}     c^{(-)}_{a,b} h^{(-)}_{a,-b}
   -   \frac{1}{2\pi v_{ab}}     \sum_{s}  c^{(s)}_{a,b}  h^{(-s)}_{a,-b}
\label{coop2}
\end{eqnarray}
where
$h^{(s)}_{a,-b} \equiv  2 f^{(s)}_{a,-b} -  \delta_{ab} c^{(s)}_{a,a}$,
and the $v_{ab}$ are defined in terms of the
Fermi velocities as $v_{ab} = (v_a + v_b)/2$ \cite{foot2}.

We have solved numerically the renormalization group equations for 
a Hubbard
model with next-to-nearest neighbor hopping
$t'=-0.2t$. Our main interest focuses on the instabilities arising
from the hot spots around the momenta $(\pi,0)$ and $(0,\pi)$.
Thus we have chosen the filling to match the Fermi level with the
position of the Van Hove singularity from the saddle points. We
have considered systems with up to $N=76$ and $152$ Fermi points,
which already demand the introduction of $\approx 5900$ independent
couplings. We have assumed that the system becomes
periodic in the transverse direction, but using twisted boundary
conditions to avoid the presence of Fermi points with vanishing
Fermi velocity at $(\pi,0)$ and $(0,\pi)$. Anyhow, in this
approach we are able to capture the physics of the hot spots by
increasing the number $N$, since the lowest Fermi velocity
(corresponding to $k_a=\pi/N$ and $k_a=\pi -\pi/N$) scales as
$2\pi (t^2-4t'^2)^{1/2}/N $. It is worth to point out that the
model remains exactly C$_{4v}$-symmetric with the mentioned choice
of filling level.

We have started the integration of the equations in the weak
coupling regime, taking a value of the on-site Hubbard repulsion
$U = t$. We recall that, when making the passage to the electron fields
in momentum space, all the couplings in the hamiltonian (\ref{hle}) get a
contribution equal to $U/N$ from the on-site interaction. In any event,
the flow is driven to a regime where some of the couplings grow large.
In particular, at some energy scale $\Lambda^{\ast} = E_c e^{-l^{\ast}}$,
a divergence is found in some of the couplings. For $N=76$, for instance,
we find that
$l^{\ast}=4.99$. As observed in Ref. \onlinecite{lbf}, the
singularity corresponds to the divergence of interactions involving
scattering between the Fermi points with smallest velocity, i.e. those
nearest to the hot spots. The rest of the couplings remain in the weak
coupling regime.


In order to characterize the instability from the large growth of the
couplings, we have also computed the scaling of the response functions
$R^{(\eta)} (\omega )$ for singlet (SS) and triplet superconducting (TS)
order parameters, as well as for charge-density-wave (CDW) and
spin-density-wave (SDW) order parameters. The superconducting order
parameters, for instance, are given by
\begin{eqnarray}
  O_a^{(SS)} & = &  \int dp
    \left( \Psi_{a \uparrow} (p) \Psi_{-a \downarrow} (-p)
       - \Psi_{a \downarrow} (p) \Psi_{-a \uparrow} (-p) \right)     \\
  O_a^{(TS)} & = &  \int dp
    \left( \Psi_{a \uparrow} (p) \Psi_{-a \downarrow} (-p)
       + \Psi_{a \downarrow} (p) \Psi_{-a \uparrow} (-p) \right)
\end{eqnarray}
These operators build response function matrices defined in the space of
Fermi points
\begin{equation}
R_{a,b}^{(\eta)} (\omega ) = -i \int_{-\infty}^{+\infty} dt e^{i\omega t}
  \langle T O_a^{(\eta)} (t) O_b^{(\eta) \dagger} (0)    \rangle
\end{equation}
with $(\eta) = (SS), (TS) $.

Due to the nontrivial structure of the Fermi line, the symmetry of the
pair wavefunction is obtained by diagonalizing the matrices
$R_{a,b}^{(SS)}$ and $R_{a,b}^{(TS)}$ in the low-energy limit, and looking
for the eigenvector corresponding to the largest eigenvalue. We recall at
this point that the derivatives with respect to the frequency of the
response functions, $\overline{R}^{(\eta)} = \partial R^{(\eta)} /
\partial \log (\omega) $, have well-defined scaling properties\cite{sol}. In
the case of the superconducting response functions, the scaling equations read:
\begin{eqnarray}
\frac{\partial \overline{R}_{a,b}^{(SS)} (\omega ) }{\partial
\log (\Lambda )}&=&\frac{1}{\pi v_{a}} \sum_{c} [(1-\delta_{a,c})
(c^{(+)}_{a,c}+ c^{(-)}_{a,c}) +
      \delta_{a,c}   (f^{(+)}_{a,-c}+ f^{(-)}_{a,-c})]
                         \overline{R}_{c,b}^{(SS)} (\omega )               \\
\frac{\partial \overline{R}_{a,b}^{(TS)} (\omega ) }{\partial
\log (\Lambda )}&=&\frac{1}{\pi v_{a}} \sum_{c} [(1-\delta_{a,c})
(c^{(+)}_{a,c}- c^{(-)}_{a,c}) +
      \delta_{a,c}   (f^{(+)}_{a,-c}- f^{(-)}_{a,-c})]
\overline{R}_{c,b}^{(TS)} (\omega ) \, .
\end{eqnarray}
Because of the symmetry of the Fermi line, the  eigenvectors of
$R_{a,b}$ transform according to the irreducible representations
of C$_{4v}$.

In the case of density-wave response functions, given a
momentum-transfer ${\bf q}$ across the Fermi line connecting the
Fermi points $(a, \overline{b})$, the only other pair connected by ${\bf q}$
is made of the inverted points $(-\overline{b}, -a)$. We may form therefore
two different operators for the same momentum ${\bf q}$
\begin{eqnarray}
  O_1^{(CDW/SDW)} ({\bf q}) & = &  \int dp
 \left( \Psi_{a \uparrow}^{\dagger} (p+q) \Psi_{\overline{b} \uparrow} (p)
       \pm \Psi_{a \downarrow}^{\dagger} (p+q)
                           \Psi_{\overline{b} \downarrow} (p) \right)     \\
  O_2^{(CDW/SDW)} ({\bf q}) & = &  \int dp
 \left( \Psi_{-\overline{b} \uparrow}^{\dagger} (p+q) \Psi_{-a \uparrow} (p)
       \pm \Psi_{-\overline{b} \downarrow}^{\dagger} (p+q)
                                \Psi_{-a \downarrow} (p) \right)
\end{eqnarray}
which give rise to two different response functions
\begin{eqnarray}
R_{1,1}^{(\eta)} ({\bf q}, \omega ) & = &
                         -i \int_{-\infty}^{+\infty} dt e^{i\omega t}
  \langle T O_1^{(\eta)} ({\bf q}, t)
                           O_1^{(\eta) \dagger} ({\bf q}, 0)    \rangle       \\
R_{1,2}^{(\eta)} ({\bf q}, \omega ) & = &
                              -i \int_{-\infty}^{+\infty} dt e^{i\omega t}
  \langle T O_1^{(\eta)} ({\bf q}, t)
                                   O_2^{(\eta) \dagger} ({\bf q}, 0)  \rangle
\end{eqnarray}

The response functions $R_{1,1}^{(\eta)} ({\bf q}, \omega )$ and
$R_{1,2}^{(\eta)} ({\bf q}, \omega )$ are actually entangled by the interactions,
as can be seen from the scaling equations
\begin{eqnarray}
\frac{\partial \overline{R}_{1,1}^{(CDW)} ({\bf q}, \omega ) }{\partial
\log (\Lambda )}&=&\frac{2}{\pi (v_{a}+v_{b})} [(c^{(+)}_{a,-\overline{b}}-2
c^{(-)}_{a,-\overline{b}})\overline{R}_{1,2}^{(CDW)} ({\bf q}, \omega  ) +
(f^{(+)}_{a,\overline{b}}-2
f^{(-)}_{a,\overline{b}})\overline{R}_{1,1}^{(CDW)} ({\bf q}, \omega  )]     \\
\frac{\partial \overline{R}_{1,2}^{(CDW)} ({\bf q}, \omega   ) }{\partial
\log (\Lambda )}&=&\frac{2}{\pi (v_{a}+v_{b})} [(c^{(+)}_{a,-\overline{b}}-2
c^{(-)}_{a,-\overline{b}})\overline{R}_{1,1}^{(CDW)} ({\bf q}, \omega   )+
(f^{(+)}_{a,\overline{b}}-2
f^{(-)}_{a,\overline{b}})\overline{R}_{1,2}^{(CDW)} ({\bf q}, \omega   )]      \\
\frac{\partial \overline{R}_{1,1}^{(SDW)} ({\bf q}, \omega   ) }{\partial
\log (\Lambda )}&=&\frac{2}{\pi (v_{a}+v_{b})}
[c^{(+)}_{a,-\overline{b}}\overline{R}_{1,2}^{(SDW)} ({\bf q}, \omega   )+
f^{(+)}_{a,\overline{b}}\overline{R}_{1,1}^{(SDW)} ({\bf q}, \omega   )]       \\
\frac{\partial \overline{R}_{1,2}^{(SDW)} ({\bf q}, \omega   ) }{\partial
\log (\Lambda )}&=&\frac{2}{\pi (v_{a}+v_{b})}
[c^{(+)}_{a,-\overline{b}}\overline{R}_{1,1}^{(SDW)} ({\bf q}, \omega   )+
f^{(+)}_{a,\overline{b}}\overline{R}_{1,2}^{(SDW)} ({\bf q}, \omega   )] \,
.
\end{eqnarray}

In the charge-density-wave and spin-density-wave sectors, the symmetry of the
dominant correlations is determined by finding the largest eigenvalue from
the diagonalization of the $2 \times 2$ blocks
\begin{eqnarray}
\left(
    \begin{array}{cc}
R^{(\eta)}_{1,1} ({\bf q}, \omega ) & R^{(\eta)}_{1,2} ({\bf q}, \omega )  \\
R^{(\eta)}_{1,2} ({\bf q}, \omega ) &  R^{(\eta)}_{1,1} ({\bf q}, \omega )
    \end{array}
  \right)
\end{eqnarray}
for $\eta = CDW, SDW $.

\begin{figure}
\includegraphics[height=5.0cm ]{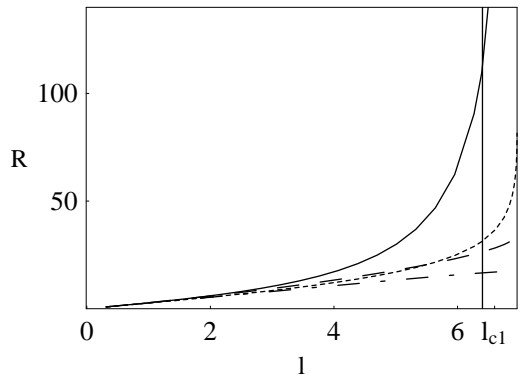}
\caption{Plot of the dominant response functions for each
different kind of order as a function of $l=-\log
(\Lambda/E_{c})$. From top to bottom, we have SS $d_{x^{2}-y^{2}}$
(solid), CDW (dotted), SDW (dashed) and TS (dotted-dashed)
response functions. The data correspond to a Hubbard
model with $N=48$, $t'=-0.2t$ and $U=t$, at the Van
Hove filling. The point where the renormalization group flow
diverges is at $l^{\ast} \approx 6.96$. The vertical line at
$l_{c1} \approx 6.40$ indicates the energy scale at which the
Fermi line starts to collapse.} 
\label{corr}
\end{figure}

\begin{figure}
\includegraphics[height=5.0cm ]{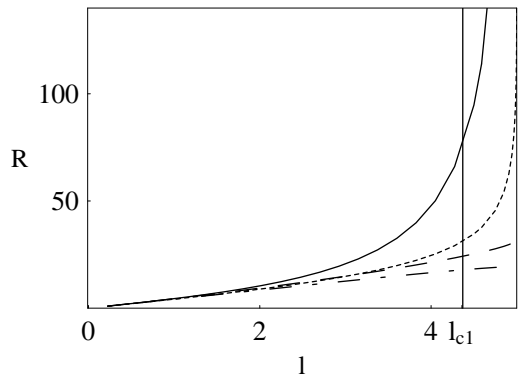}
\caption{Similar plot as in Fig. \ref{corr} for $N=76$, with $l^{\ast} \approx
4.99$ and $l_{c1} \approx 4.36$.} 
\label{corr2}
\end{figure}

We have plotted in Figs. \ref{corr} and \ref{corr2} the evolution of 
the response functions for arrays with $N=48$ and 76, respectively, 
taking as a representative 
the largest eigenvalue for each kind of order, i.e. SS, TS, CDW and SDW.
The dominant instability is found in the superconducting channel with
$d_{x^2-y^2}$ symmetry. The corresponding eigenvector of the matrix
$R_{a,b}^{(SS)}$ represents the amplitude of the Cooper pair wave function
along the Fermi line, which turns out to be peaked (with opposite signs)
at the two hot spots around $(\pi , 0)$ and $(0, \pi)$, and vanishing
elsewhere. This result is consistent with the dominance of the $d$-wave
correlations observed in $N$-leg Hubbard ladders for small values of
$N$.

\

At this point, we can go beyond the renormalization group solution of the 
model, by incorporating the nonperturbative methods used in the previous section
for the analysis of the quasiparticle properties. Our approach consists in
computing the string of 
anomalous scaling dimensions $\{ \gamma_a \}$ with the effective couplings
at each renormalization group step.
To this end, we use the couplings calculated by solving the renormalization
group equations to diagonalize the Hamiltonian in Eq. (\ref{hll})
for $0<l<l^{\ast}$. On technical grounds, this amounts to
bosonize the electron system at different values of $\Lambda $.
We end up in this way with an anomalous electron dimension $\gamma_a (l)$
that depends on the energy scale. The
numerical computation of $\gamma_a (l)$ reveals that, at a scale 
$l_{c1} < l^{\ast}$, the anomalous dimension for the Fermi points
nearest to $(\pi , 0)$ and $(0, \pi )$ diverges (namely for $k_a= \pi/N$ and
$k_a = \pi -\pi/N$), while $\gamma_a(l_{c1})$ remains small for the rest
of the electron fields. 

The divergence of the anomalous dimension $\gamma_a (l)$ near $(\pi , 0)$ and 
$(0, \pi )$ can be traced back to the singular behavior of the change of 
variables (\ref{cv1})-(\ref{cv2}) at the point $l = l_{c1}$. The corresponding 
operators $B^{\dagger }_{a} (q)$ and $B_{a} (q)$ have projections 
$\tilde{s}_{ab}, \tilde{s}_{\overline{a} \overline{b}}, \tilde{t}_{a \overline{b}}$ 
and $\tilde{t}_{\overline{a} b}$ that diverge for particular components
$\widetilde{B}^{\dagger }_b (q)$ and $\widetilde{B}_b (q)$. These are made of
a $C_{4v}$-symmetric combination of operators peaked at the points nearest 
to $(\pi , 0)$ and $(0, \pi )$. When rewritten in terms of the density 
operators $B^{\dagger }_{a} (q) + B_{\overline{a}} (-q)$, the change of variables 
leads however to a projection $\tilde{s}_{ab} + \tilde{t}_{\overline{a} b}$  
$(= \tilde{s}_{\overline{a} \overline{b}} + \tilde{t}_{a \overline{b}})$ that 
vanishes at $l = l_{c1}$ for the mentioned component. This shows that 
the situation is similar to that found in a 1D system with two subbands, where 
under certain conditions the Luttinger liquid parameter $K_+$ in the channel 
of the total charge density may vanish at a certain point of
the renormalization group flow\cite{jv}. The correspondence is such that 
$\tilde{s}_{ab}  \sim  (\sqrt{K_+} + 1/\sqrt{K_+})/2$ and 
$\tilde{t}_{\overline{a} b}   \sim  (\sqrt{K_+} - 1/\sqrt{K_+})/2$.
Specifically, in the model of two subbands with equal Fermi velocity
$v_a = v_b = v_F$,
\begin{equation}
K_+ = \sqrt{ \frac{\pi v_F + (f^{(+)}_{a,a} + f^{(+)}_{a,b}) 
                - (f^{(+)}_{a,\overline{a}} + f^{(+)}_{a,\overline{b}})}
               {\pi v_F + (f^{(+)}_{a,a} + f^{(+)}_{a,b}) 
                + (f^{(+)}_{a,\overline{a}} + f^{(+)}_{a,\overline{b}})}  }
\end{equation}
We have checked that, as well as in the system of two subbands, the vanishing
of $\tilde{s}_{ab} + \tilde{t}_{\overline{a} b}$ takes place in our 
model when the interband interactions $f^{(+)}_{a,\overline{b}}$ between the 
points nearest to $(\pi , 0)$ and $(0, \pi )$ grow large
enough upon renormalization to drive to zero the corresponding Luttinger 
liquid parameter.

We have then identified the origin of the singular behavior in the change of
variables (\ref{cv1})-(\ref{cv2}). This takes place when the couplings
$f^{(+)}_{a,\overline{b}}$
become of the order of the Fermi velocity $v_a$, that is, when they
are leaving the weak-coupling regime. Thus, the divergence in the anomalous 
scaling dimension at $l = l_{c1}$ is found before entering the regime of 
large growth of the renormalized couplings, and it may be considered a
more reliable result than the predictions obtained from the large growth 
of the response functions. 

The above discussion serves also to clarify the physical meaning of the 
divergence of $\gamma_a $. This is the reflection of a local instability
in the string of Fermi points, since at this stage it only affects the points
nearest to $(\pi , 0)$ and $(0, \pi )$. We can first analyze the 
effect of the divergence of $\gamma_a $ by recalling that this anomalous
dimension dictates the behavior of the observables defined from the electron
propagator. In particular, the density of states, that we define here as a 
quantity $n_a (\varepsilon )$ depending on the Fermi point, is obtained by
taking the Fourier transform of the time-dependence of the electron
propagator
\begin{equation}
\langle \psi^{\dagger }_{a \sigma} (x,t)  \psi_{a \sigma} (x,t') \rangle
       \sim  \frac{1}{| t - t' |^{1 + \gamma_a }}
\end{equation}
Thus, we get the power-law behavior
\begin{equation}
n_a (\varepsilon ) \sim \varepsilon^{\gamma_a}
\end{equation}
where the energy is measured in units of the high-energy cutoff.
Very large values of the exponent $\gamma_a $ indicate a very strong suppression 
of the electron quasiparticles, as the shape of the density of states becomes 
almost flat at low energies. We see that, at the point where $\gamma_a $ 
diverges, a gap opens effectively in the spectrum of fermion excitations, since 
the density of states vanishes then over a range of low energies.

\begin{figure}
\includegraphics[height=5.0cm ]{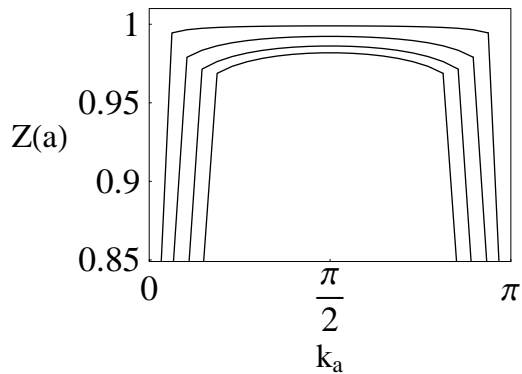}
\caption{Angle-resolved quasiparticle weight $Z_a$ as a function of the
transverse momentum $k_a$ for the first 4 iterations of the combined
renormalization group and bosonization approaches for an array with $N=48$. 
From top to bottom,
$Z_a$ is shown at the points $l_{c1},l_{c2},l_{c3},l_{c4}$ where a new set of
Fermi points disappears from the spectrum. The parameters are the same as in
Fig. \ref{corr}.}
\label{zeta}
\end{figure}

The opening of a gap near $(\pi , 0)$ and $(0, \pi )$ in the string of Fermi
points can be studied more precisely in our renormalization group framework.
For this purpose we deal with the angle-resolved quasiparticle
weight $Z_a$, which is related to the anomalous dimension $\gamma_a $ through
the scaling equation
\begin{equation}
\frac{1}{Z_a (l)}
\frac{\partial Z_a (l) }{ \partial l} = -\gamma_{a}(l) \, .
\label{zetaeq}
\end{equation}
We have integrated numerically Eq. (\ref{zetaeq}) for every $a$ in the
range $0 < l < l_{c1}$. The result is shown in Fig. \ref{zeta} (top line)
for $N=48$.
We observe that the quasiparticle weight vanishes at $l = l_{c1}$ for the Fermi 
points nearest to $(\pi , 0)$ and $(0, \pi )$, while in the rest of the Fermi 
line the renormalization is quite mild. We remark that the vanishing of $Z_a$ 
indicates the disappearance of fermion excitations at the Fermi point $a$
below an energy given by $\Lambda_1=E_c e ^{-l_{c1}}$. In this respect, the
physical effect is different than in a Luttinger liquid where the anomalous
dimension of the electron is independent of the energy scale. A constant value
of $\gamma_a $ implies a quasiparticle weight $Z_a \sim \exp (-l \gamma_a )$,
vanishing only at zero energy. In our case, however, the quasiparticle weight
vanishes at an energy $\Lambda_1 \neq 0$, implying the opening of
a gap below that energy scale at the points near $(\pi,0)$ and $(0,\pi)$.

It is interesting to follow the trend of the rest of Fermi points
by further lowering the energy scale. This is achieved by solving
again the renormalization group equations for the set of Fermi
points obtained by excluding the decoupled ones. For $N=48$ we
remain with $96-8=88$ Fermi points, and the new initial conditions 
are set by evaluating the corresponding couplings at $l_{c1}$, 
as obtained from the previous renormalization group flow. According 
to the above discussion, all of them are still in the weak coupling 
regime, making the calculation well-posed. The new renormalization 
group flow is obtained numerically starting from $l_{c1}$.
Again an instability is observed, due to the growth of the
interactions involving scattering between Fermi points with
smaller Fermi velocity (i.e. at $k_a=3\pi/N$ and $k_a = \pi -
3\pi/N$). Following the same procedure shown above,  the new
anomalous dimensions $\gamma_a$ and the quasiparticle weights
$Z_a$ are computed. Also in this case, the $\gamma_a $ for the
Fermi points nearest to the hot spots diverge at $l_{c2} >
l_{c1}$. The new quasiparticle weights are
obtained by solving Eq. (\ref{zetaeq}) in the range
($l_{c1},l_{c2}$), with initial conditions dictated by the
previous integration of the same equations. In Fig. \ref{zeta}
(second line from top), we see that $Z_a$ vanishes again only for
the points with smaller Fermi velocity. We observe that its
renormalization is now more pronounced at different angles with
respect to the previous stage.

This procedure can be iterated, giving rise to the progressive
destruction of the Fermi line, from $(\pi , 0)$ and $(0, \pi )$ towards
the diagonals of the Brillouin zone. It should be noted that the
energy scale $\Lambda_n = E_c e^{-l_{cn}}$ at which the $n$-th Fermi
point disappears may become very small.
In Fig. \ref{zeta} we have plotted the angular distribution of $Z_a$
for the first 4 iterations, corresponding to the destruction of the
Fermi points having transverse momentum $\pm (2n + 1)\pi/48$ and
$\pm \pi \mp (2n + 1)\pi/48$ , with $n=0,1,2,3$. We observe that,
by increasing the number of iterations, the renormalization of $Z_a$
away from the hot spots becomes more and more pronounced.
The plot is actually reminiscent of the results for the quasiparticle
renormalization shown in Ref. \onlinecite{zan}.
There is a clear indication that, while the collapse occurs on
finite arcs around $(\pi,0)$ and $(0,\pi)$, the rest of the Fermi
line results renormalized by the interactions.

\begin{figure}
\includegraphics[height=5.0cm ]{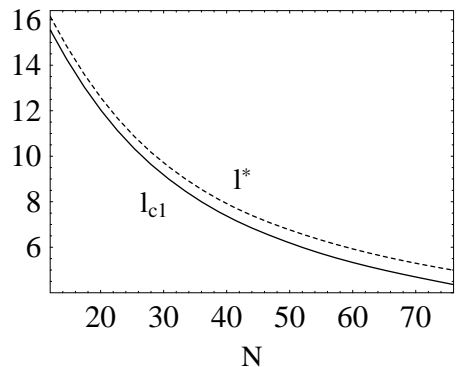}
\caption{Plot of the critical scales  $l^{\ast}$ (dotted line) and
$l_{c1}$ (full line) as a function of $N$. The
parameters are the same as in Fig. \ref{corr}.} 
\label{scale}
\end{figure}

The comparison of the plots in Figs. \ref{corr} and \ref{corr2}
gives a fair idea about the dependence of the above results 
on the size of the array of chains. The $d$-wave
superconducting correlations turn out to be always dominant (see
Fig. \ref{corr2}), but the collapse of the Fermi line at the
hot spots prevents them from growing large in any case. In Fig.
\ref{scale} we have represented 
the trend of the critical scales  $l^{\ast}$ and
$l_{c1}$ as a function of $N$. It appears that $l^{\ast}$
approaches an asymptotic value in the large-$N$ limit,
while the difference between $l^{\ast}$ and $l_{c1}$
does not vanish by increasing the size of the array. This
suggests that the development of the pseudogap should survive in the
2D limit, ending up with a paradigm of electron liquid in
which the Fermi surface is progressively destroyed as the system
is probed at lower energy scales.

\section{Conclusion}

In this paper we have studied the low-energy electronic instabilities in
a 2D Hubbard model formed by a large array of coupled
chains. Within a low-energy cutoff $E_c$ about the Fermi level, the different
interactions can be classified into a reduced number of channels\cite{lbf},
which coincide essentially with those obtained from the analysis of the
kinematics for a 2D Fermi surface\cite{shankar}. In the model with bare
on-site repulsion $U$, the different couplings
in momentum space have a strength of the order of $U/N$, which makes
appropriate the use of renormalization group methods to determine the
scaling of the couplings as the cutoff is reduced towards the Fermi level.

Complementing this approach, we have also applied bosonization methods
to describe the dressing of the electron fields at the different Fermi
points. Thus, we have computed the string of 
anomalous scaling dimensions $\{ \gamma_a \}$
with the effective couplings at each renormalization group step. This
combines the power of the renormalization group and
the bosonization methods, allowing to characterize the points where the
metallic phase may break down. In this way, we have
been able to discern the competition between the tendency to the
formation of ordered phases in the 2D electron liquid and the charge
instabilities that modify the shape of the Fermi line.

Our main interest has been focused on the instabilities that appear
when the Fermi level is close to the Van Hove singularity arising from
the saddle points at $(\pi ,0)$ and $(0, \pi )$. It is well-known that
the singularity in the density of states leads to strong correlations
that signal the tendency towards a spin-density-wave phase or $d$-wave
superconductivity at low energies. This has been studied in the past
using mainly renormalization group methods. An important drawback in
this approach comes however from the unconventional form of the 2D
perturbative expansion, with terms that diverge at low energies as the
square of $\log (\varepsilon ) $ and which are not therefore properly
renormalizable. This may cast some doubts about the complete
predictability of the renormalization group to discern the competition
between low-energy instabilities in the 2D model.

Alternatively, it has been pointed out that the renormalization
of the quasiparticle properties has to be quite significant near the
Van Hove singularity, and that this may lead to a drastic softening of
the electron correlations\cite{zan}. In our approach, we have recovered
some of the trends found within the conventional renormalization group
schemes, like the strong renormalization of the dispersion relation
around the saddle points\cite{nucl,kat2} or the growth of the $d$-wave
superconducting correlations\cite{dwave,eur,schu,jap,metz,prl,hon,binz,kat}.
However, our main finding is that, at the Van Hove filling, the progressive
destruction of the Fermi line takes place around the saddle points before
any correlation may diverge in the low-energy limit. This is a sensible
result in our model, as the 1D renormalization group
approach is well-defined, while the analysis of the breakdown of the metallic
phase is carried out with a nonperturbative bosonization approach.

This study may have also implications for the understanding of the pseudogap
phase in the copper-oxide superconductors. The loss of fermion excitations 
takes place in our model as the electron fields are attenuated
due to the dressing from the renormalized interactions at the hot spots
around $(\pi , 0)$ and $(0, \pi )$. This happens as the charge 
stiffness goes to zero at some
of the Fermi points in the low-energy limit. We stress that this behavior
marks the approach of a critical point, characterized by the vanishing
of some of the Luttinger liquid parameters encoded in the expression
(\ref{anmd}) for the anomalous dimension\cite{jv,foot3}. The critical point
is reached as some of the effective couplings leave the weak-coupling
regime, so our description of the pseudogap does not rely on strong-coupling
features to account for the destruction of the Fermi line. In future
studies, it would be interesting to investigate the extrapolation of
the present results to systems with larger number of chains.
It seems plausible that the
destruction of the Fermi line we have described should persist for
increasing number of chains, as it relies on a critical behavior that
is a robust property of the bosonization of the system in the
low-energy limit.

\section*{Acknowledgements}
The financial support of the Ministerio de Educaci\'on y Ciencia
(Spain) through grant FIS2005-05478-C02-02 is gratefully
acknowledged. E. P. was also supported by Consorzio Nazionale
Interuniversitario per le Scienze Fisiche della Materia.



\begin{thebibliography}{99}
\bibitem{foot}
This set of low-energy couplings becomes completely equivalent to that
introduced in Ref. \onlinecite{lbf}, after assuming the spin-independence
of the interactions in every channel.
\bibitem{note}
For even $N$, 
the set of couplings in the hamiltonian (\ref{hle}) has to be in
principle completed with a number of vertical Umklapp processes 
(direct and exchange), in which the transverse momentum $k_a$ is
conserved modulus $2 \pi$ (see Ref. \onlinecite{lbf}). We have checked 
by numerical inspection that their couplings remain small under
the renormalization group flow and that they do not affect the flow of 
the rest of couplings so, for the sake of simplicity, we have not 
included them explicitly in the general discussion.
\bibitem{foot2}
An equivalent set of equations for a different choice of independent
couplings can be found in Ref. \onlinecite{lbf}.
\end{thebibliography}
\end{document}